\documentstyle[12pt]{article}

\begin{document}
\begin{flushright}
February 11, 2016
\end{flushright}

\makeatletter
\@addtoreset{equation}{section}
\def\theequation{\thesection.\arabic{equation}}
\makeatother

\vskip 0.5 truecm
\vskip 0.5 truecm

\begin{center}
{\Large{\bf  The gradient flow in $\lambda\phi^{4}$ theory}}
\end{center}
\vskip .5 truecm
\centerline{\bf Kazuo Fujikawa}
\vskip .4 truecm
\centerline{\it Quantum Hadron Physics Laboratory,}
\centerline{\it RIKEN Nishina Center, Wako 351-0198, Japan}
\vskip 0.5 truecm

\begin{abstract}
A gradient flow equation for $\lambda\phi^{4}$ theory in $D=4$ is formulated. In this scheme the gradient flow equation is written in terms of the renormalized probe variable $\Phi(t,x)$ and renormalized parameters $m^{2}$ and $\lambda$ in a manner analogous to the higher derivative regularization. No extra divergence is induced in the interaction of the probe variable $\Phi(t,x)$  and the 4-dimensional dynamical variable $\phi(x)$ which is defined in renormalized perturbation theory. The finiteness to all orders in perturbation theory is established  by power counting argument in the context of $D+1$ dimensional field theory. This illustrates that one can formulate the gradient flow for the simple but important $\lambda\phi^{4}$ theory in addition to the well-known Yang-Mills flow, and it shows the generality of the gradient flow for a wider class of field theory.
\end{abstract}



\section{Introduction}
It has been noted recently that the simplest $\lambda\phi^{4}$ theory  in $D=4$ is not incorporated in the framework of the gradient flow in the sense that the flow drives the fields to configurations corresponding to minima of the action~\cite{monahan1, ehret, monahan}. This is in sharp contrast to the case of pure Yang-Mills field theory~\cite{luscher1, luscher2} with a slight modification in the case of QCD~\cite{luscher3}. See also a recent review~\cite{ramos} and earlier related works~\cite{morningstar, neuberger}. In the general context of smearing in field theory, one should include the stout-smearing~\cite{morningstar} as well as the gradient flow in a narrow sense which drives the fields to minima of the action~\cite{luscher1, luscher2}. For example, the smeared operator product
expansion~\cite{monahan} and the analysis of energy-momentum tensor~\cite{ehret} both of which are based on the simple Gaussian smearing  in $\lambda\phi^{4}$ theory, in addition to the smearing of fermions in QCD~\cite{luscher3}, are counted as interesting applications of the idea of smearing.
In the present paper, however, we use the term "gradient flow" in a rather narrow sense which drives the fields to minima of the action.

If one attempts at the gradient flow in $\lambda\phi^{4}$ theory by incorporating interactions in the flow time
evolution, as is done in pure Yang-Mills theory, the wave-function, mass and coupling constant renormalization factors need to satisfy certain relations if one asks the gradient flow equation to be consistently defined in terms of {\em bare quantities}.  But the actual $\lambda\phi^{4}$ theory in $D=4$ does not satisfy these conditions. This formulation of $\lambda\phi^{4}$ theory would thus 
remove the renormalization property of the gradient flow that proves useful; renormalized correlation
functions would no longer be guaranteed to remain finite at non-zero flow time. It has been widely believed that this difference stems from the internal symmetries of pure Yang-Mills theory that has no analogue in scalar $\lambda\phi^{4}$ theory in $D=4$; in pure Yang-Mills theory, it is gauge
invariance, manifested through appropriate BRST symmetries, that ensures no new counterterms generated by the gradient flow~\cite{monahan1}. We want to show in the present paper that the gradient flow of scalar $\lambda\phi^{4}$ theory in $D=4$ can also be consistently defined if suitably formulated.

A way to avoid the complications in $\lambda\phi^{4}$ theory in $D=4$ mentioned above may be to write the gradient flow equation in terms of renormalized quantities~\cite{aoki}. 
Even in this case, it has been recognized that divergences which cannot be cancelled by any counter term  appear in the flow equation in $D=4$~\cite{suzuki}. To be explicit, one may start with the simplest flow equation in the Euclidean metric $g_{\mu\nu}=(1,1,1,1)$,
\begin{eqnarray}
\dot{\Phi}(t,x)=\Box\Phi(t,x)-m^{2}\Phi(t,x)-\lambda\Phi^{3}(t,x)
\end{eqnarray}
with $\dot{\Phi}=\frac{d}{dt}\Phi$ for $t\geq 0$ and renormalized mass $m^{2}$ and coupling constant $\lambda$, and the initial condition
\begin{eqnarray}
\Phi(0,x)=\phi(x)
\end{eqnarray}
where $\phi(x)$ stands for the renormalized field.
When one expands
\begin{eqnarray}
\Phi(t,x)=\varphi_{0}(t,x)+\lambda\varphi_{1}(t,x)+\lambda^{2}\varphi_{2}(t,x)+ .... ,
\end{eqnarray}
one has
\begin{eqnarray}
\dot{\varphi}_{0}(t,x)+(-\Box+m^{2})\varphi_{0}(t,x)&=&0,\nonumber\\
\dot{\varphi}_{1}(t,x)+(-\Box+m^{2})\varphi_{1}(t,x)&=&-\varphi_{0}(t,x)^{3},\nonumber\\
&.......&
\end{eqnarray}
with
\begin{eqnarray}
\varphi_{0}(0,x)=\phi(x), \ \ \ \varphi_{1}(0,x)=0.
\end{eqnarray}
Thus we have
\begin{eqnarray}
\varphi_{0}(t,x)&=&e^{-(-\Box+m^{2})t}\phi(x),\nonumber\\
&=&\int d^{4}y K(t,x-y)\phi(y)
\end{eqnarray}
with
\begin{eqnarray}
K(t,x-y)&=&\langle x|e^{-(-\Box+m^{2})t}|y\rangle\nonumber\\
&=&\int \frac{d^{4}k}{(2\pi)^{4}}e^{-k^{2}t+ik(x-y)-m^{2}t}\nonumber\\
&=&\frac{1}{16\pi^{2}}\frac{1}{t^{2}}\exp{[-\frac{(x-y)^{2}}{4t}-m^{2}t]}.
\end{eqnarray}
We also have
\begin{eqnarray}
\varphi_{1}(t,x)
&=&-\int_{0}^{t}ds\int d^{4}y  K(t-s,x-y)\\
&&\times[ \int d^{4}z_{1}d^{4}z_{2}d^{4}z_{3}K(s,y-z_{1})K(s,y-z_{2})K(s,y-z_{3})\phi(z_{1})\phi(z_{2})\phi(z_{3})].\nonumber
\end{eqnarray}
When one contracts two fields $\phi(z_{1})$ and $\phi(z_{2})$, for example, to $\langle \phi(z_{1})\phi(z_{2})\rangle$ to
examine an (open) tadpole-type diagram in $\varphi_{1}(t,x)$, namely, a loop diagram of $\phi(x)$ detached from the main part of the Green's functions in  $\lambda\phi^{4}$ theory, one obtains the factor 
\begin{eqnarray}
f(t)=\int\frac{d^{4}k}{(2\pi)^{4}}\frac{1-e^{-2(k^{2}+m^{2})t}}{2(k^{2}+m^{2})}\frac{1}{k^{2}+m^{2}},
\end{eqnarray}
which diverges even for $t>0$~\cite{suzuki}.  This divergence is not cancelled by any counter term, and this shows the failure of the renormalized gradient flow equation (1.1) to define the finite probe variable. (See also eq.(2.22) below.)

\section{A proposal}
We here propose a way to make the renormalized gradient flow equation for the simplest $\lambda\phi^{4}$ theory in $D=4$ consistent.

We start with the Euclidean field theory
\begin{eqnarray}
\int {\cal D}\phi_{0} \exp\{ \int d^{4}x{\cal L}\}
\end{eqnarray}
with
\begin{eqnarray}
{\cal L}&=&-\frac{1}{2}\partial_{\mu}\phi_{0}(x)\partial_{\mu}\phi_{0}(x)-\frac{1}{2}m_{0}^{2}\phi_{0}^{2}(x)-\frac{1}{4}\lambda_{0}\phi_{0}^{4}(x)\nonumber\\
&=&-\frac{1}{2}\partial_{\mu}\phi(x)\partial_{\mu}\phi(x)-\frac{1}{2}m^{2}\phi^{2}(x)-\frac{1}{4}\lambda\phi^{4}(x) +{\cal L}_{counter},
\end{eqnarray}
where $m^{2}$, $\lambda$ and $\phi$ all stand for the finite renormalized quantities, and ${\cal L}_{counter}$ contains all the counter terms that render all the Green's functions $\langle  T^{\star}\phi(x_{1})\phi(x_{2})\phi(x_{3}) .....\rangle$ which are defined by 
\begin{eqnarray}
\langle T^{\star}\phi(x_{1})\phi(x_{2})\phi(x_{3}) ....\rangle=\int {\cal D}\phi \left(\phi(x_{1})\phi(x_{2})\phi(x_{3}) ....\right)\exp\{ \int d^{4}x{\cal L}\}
\end{eqnarray}
finite.
The Euclidean metric convention is $g_{\mu\nu}=(1,1,1,1)$.
The operator equation of motion is then
\begin{eqnarray}
\langle -\Box\phi(x)+m^{2}\phi(x)+\lambda\phi^{3}(x)-\frac{\delta{\cal L}_{counter}}{\delta \phi(x)}\rangle =0.
\end{eqnarray}
The flow equation we propose is defined by
\begin{eqnarray}
\dot{\Phi}(t,x)=-(-\Box+m^{2})^{2}\Phi(t,x)-\lambda(-\Box+m^{2})\Phi^{3}(t,x)
\end{eqnarray}
with $t\geq 0$ and 
\begin{eqnarray}
\Phi(0,x)=\phi(x).
\end{eqnarray}
This prescription is allowed since the gradient flow does not modify the basic dynamics of the starting field theory, and it does not alter the possible asymptotic equation  $(-\Box+m^{2})\Phi(t,x)+\lambda\Phi^{3}(t,x)=0$
arising from $\dot{\Phi}(t,x)=0$ for $t\rightarrow\infty$ in Euclidean theory. 
It should be noted that the "time" $t$ in (2.5) has mass dimensions $[t]=[M]^{-4}$ in contrast to the "time" $t$ in (1.1) which has $[t]=[M]^{-2}$, although we use the same notation for simplicity. 

It is suggestive to write our proposed flow equation in the form
\begin{eqnarray}
\dot{\Phi}(t,x)
&=&-\hat{D}_{x}[(\hat{D}_{x}+\lambda \Phi^{2}(t,x))\Phi(t,x)]
\end{eqnarray}
with 
\begin{eqnarray}
\hat{D}_{x}=-\Box+m^{2},
\end{eqnarray}
in comparison with Yang-Mills theory.

We then expand
\begin{eqnarray}
\Phi(t,x)=\varphi_{0}(t,x)+\lambda \varphi_{1}(t,x)+\lambda^{2}\varphi_{2}(t,x)+ .... ,
\end{eqnarray}
and we have
\begin{eqnarray}
\dot{\varphi}_{0}(t,x)+(-\Box+m^{2})^{2}\varphi_{0}(t,x)&=&0,\nonumber\\
\dot{\varphi}_{1}(t,x)+(-\Box+m^{2})^{2}\varphi_{1}(t,x)&=&
-(-\Box+m^{2})\varphi_{0}(t,x)^{3},\nonumber\\
\dot{\varphi}_{2}(t,x)+(-\Box+m^{2})^{2}\varphi_{2}(t,x)&=&
-(-\Box+m^{2})3\varphi_{1}(t,x)\varphi_{0}(t,x)^{2},\nonumber\\
\dot{\varphi}_{3}(t,x)+(-\Box+m^{2})^{2}\varphi_{3}(t,x)&=&
-(-\Box+m^{2})\nonumber\\
&&\times[3\varphi_{2}(t,x)\varphi_{0}(t,x)^{2}+3\varphi_{0}(t,x)\varphi_{1}(t,x)^{2}],\nonumber\\
&& ..... ,
\end{eqnarray}
with the initial conditions,
\begin{eqnarray}
\varphi_{0}(0,x)=\phi(x), \ \ \ \varphi_{1}(0,x)=0,\ \ \ \varphi_{2}(0,x)=0, ... .
\end{eqnarray}
Thus we have
\begin{eqnarray}
\varphi_{0}(t,x)&=&e^{-(-\Box+m^{2})^{2}t}\phi(x),\nonumber\\
&=&\int d^{4}y K(t,x-y)\phi(y)
\end{eqnarray}
with
\begin{eqnarray}
K(t,x-y)&=&\langle x|e^{-(-\Box+m^{2})^{2}t}|y\rangle\nonumber\\
&=&\int \frac{d^{4}k}{(2\pi)^{4}}e^{-(k^{2}+m^{2})^{2}t+ik(x-y)}
\end{eqnarray}
whose explicit form is not obtained but it defines  smearing  with respect $x$ and $y$ and 
\begin{eqnarray}
K(0,x-y)&=&\delta(x-y).
\end{eqnarray}
We thus have
\begin{eqnarray}
\varphi_{1}(t,x)&=&-\int_{0}^{t}ds\int d^{4}y K(t-s,x-y)(-\Box+m^{2})_{y}\varphi_{0}(s,y)^{3}\nonumber\\
&=&-(-\Box+m^{2})_{x}\int_{0}^{t}ds\int d^{4}y K(t-s,x-y)\varphi_{0}(s,y)^{3}\nonumber\\
\varphi_{2}(t,x)&=&-\int_{0}^{t}ds\int d^{4}y K(t-s,x-y)(-\Box+m^{2})_{y}[3\varphi_{1}(s,y)\varphi_{0}(s,y)^{2}]\nonumber\\
&=&-(-\Box+m^{2})_{x}\int_{0}^{t}ds\int d^{4}y K(t-s,x-y)[3\varphi_{1}(s,y)\varphi_{0}(s,y)^{2}]\nonumber\\
& ...... &
\end{eqnarray}
after partial integrations. It is significant that a factor $(-\Box+m^{2})_{x}$ is taken outside the main body of the expression, and the main body with the kernel $K(t-s,x-y)$ is specified by the higher derivative operator; this is the basic mechanism how  the present scheme works as is shown later.

To be explicit,  we have, for example,  
\begin{eqnarray}
\varphi_{1}(t,x)
&=&-(-\Box+m^{2})_{x}\int_{0}^{t}ds\int d^{4}y  K(t-s,x-y)\\
&\times&\int d^{4}z_{1}d^{4}z_{2}d^{4}z_{3}K(s,y-z_{1})K(s,y-z_{2})K(s,y-z_{3})\phi(z_{1})\phi(z_{2})\phi(z_{3}).\nonumber
\end{eqnarray}
When this operator is inserted into the ordinary Green's functions, one obtains the 
correlation functions such as $\langle T^{\star}\phi(z_{1})\phi(z_{2})\phi(z_{3})\phi(y_{1})\phi(y_{2})......\rangle$. In this paper we deal with perturbative expansions, and thus the field variable $\phi(x)$ may be regarded to be defined in the interaction picture. We often discard the symbol $T^{\star}$ in the present Euclidean theory. To define the ordinary Green's functions for $\lambda\phi^{4}$ theory in perturbation theory, we implicitly assume the dimensional regularization~\cite{thooft, collins} or a more conventional regularization which reproduces the results of dimensional regularization~\cite{fujikawa}.

\subsection{ Some sample calculations}

To confirm that our proposal is working in lower order processes, we perform some sample calculations. 
We start with contracting two fields $\phi(z_{1})$ and $\phi(z_{2})$, for example,
\begin{eqnarray}
\langle \phi(z_{1})\phi(z_{2})\rangle,
\end{eqnarray}
to estimate corrections to $\varphi_{1}(t,x)$ arising from  an (open) tadpole-type diagram in perturbation theory. 
The expression (2.16) for $\varphi_{1}(t,x)$  is written as 
\begin{eqnarray}
&&\int [dp][-(p^{2}+m^{2})e^{ipx}]\int [dp_{1}][dp_{2}][dp_{3}][\frac{e^{-(d(p_{1})+d(p_{2})+d(p_{3}))t}-e^{-d(p)t}}{d(p)-d(p_{1})-d(p_{2})-d(p_{3})}]\nonumber\\
&&\times(2\pi)^{4}\delta(p-p_{1}-p_{2}-p_{3})\phi(p_{1})\phi(p_{2})\phi(p_{3})
\end{eqnarray}
where we defined 
\begin{eqnarray}
d(p)\equiv [p^{2}+m^{2}]^{2}, \ \ \ \ [dp]\equiv \frac{d^{4}p}{(2\pi)^{4}}.
\end{eqnarray}
By noting the free propagator
\begin{eqnarray}
\langle \phi(p_{1})\phi(p_{2})\rangle=\int [dk](2\pi)^{4}\delta(p_{1}+p_{2})(2\pi)^{4}\delta(p_{1}-k)\frac{1}{k^{2}+m^{2}}
\end{eqnarray}
we have for (2.18)
\begin{eqnarray}
&&\int [dp][dk][-(p^{2}+m^{2})e^{ipx}\phi(p)]
[\frac{e^{-(d(k)+d(k)+d(p))t}-e^{-d(p)t}}{-d(k)-d(k)}]\nonumber\\
&&\times\frac{1}{k^{2}+m^{2}}\nonumber\\
&&=-(-\Box+m^{2})\varphi_{0}(t,x)f(t)
\end{eqnarray}
where
\begin{eqnarray}
f(t)=\int [dk]\frac{1-e^{-2d(k)t}}{2d(k)}\frac{1}{k^{2}+m^{2}}
\end{eqnarray}
which is finite and satisfies 
\begin{eqnarray}
&&f(0)=0,\nonumber\\
&&f(\infty)=\frac{1}{2}\int [dk]\frac{1}{(k^{2}+m^{2})^{3}}. 
\end{eqnarray}
This $f(t)$ may be compared to the result in (1.9).
After the one-loop correction we described, we have 
\begin{eqnarray}
\varphi_{1}(t,x)
&=&-3(-\Box+m^{2})\varphi_{0}(t,x)f(t)
\end{eqnarray}
by taking into account 3 ways to contract.

It is interesting to examine what happens if one uses this corrected result in 
$\varphi_{2}(t,x)$.
We then have 
\begin{eqnarray}
\varphi_{2}(t,x)
&=&9(-\Box+m^{2})_{x}\int_{0}^{t}ds\int d^{4}y K(t-s,x-y)\nonumber\\
&&\times[(-\Box_{y}+m^{2})\varphi_{0}(s,y)f(s)]\varphi_{0}(s,y)^{2}
\end{eqnarray}
and we examine
\begin{eqnarray}
\int_{0}^{t}ds\int d^{4}y K(t-s,x-y)[(-\Box_{y}+m^{2})\varphi_{0}(s,y)f(s)]\varphi_{0}(s,y)^{2}.
\end{eqnarray}
The contraction of the last factor $\varphi_{0}(s,y)^{2}$, namely, $\langle \varphi_{0}(s,y)^{2}\rangle$, gives a result similar to the above, and the convergence  is in fact improved by the factor $f(s)$ that suppresses the contribution near $s=0$, which causes the possible divergence, since $f(0)=0$. 
The crucial factor in this analysis is, using the explicit formula for $f(s)$ in (2.22),
\begin{eqnarray}
g(t)&\equiv&\int [dk^{\prime}]\int_{0}^{t}ds \frac{e^{-2d(k^{\prime})s}}{(k^{\prime})^{2}+m^{2}}\int [dk]\frac{1-e^{-2d(k)s}}{2d(k)}\frac{1}{k^{2}+m^{2}}
\end{eqnarray}
The evaluation of $g(t)$ is given in (A.4) in Appendix where it is shown to be convergent and $g(0)=0$. We thus have 
\begin{eqnarray}
\varphi_{2}(t,x)=9[(-\Box+m^{2})^{2}\varphi_{0}(t,x)]g(t).
\end{eqnarray}
When one studies the contraction of $[(-\Box_{y}+m^{2})\varphi_{0}(s,y)]\varphi_{0}(s,y)$ in (2.26), one examines 
\begin{eqnarray}
h(t)&\equiv&\int_{0}^{t}ds\int [dk]e^{-2d(k)s}f(s)
\end{eqnarray}
which is confirmed to be well-convergent. See (A.7) in Appendix. The convergence property is about the same as $f(t)$ in (2.22), and 
$h(0)=0$. 
The quantity in  (2.25) for this contraction is given by 
\begin{eqnarray}
\varphi_{2}(t,x)
&=&18(-\Box+m^{2})\varphi_{0}(t,x)h(t)
\end{eqnarray}
by taking into account two possible ways of contraction.

The present analysis thus implies that our proposed flow equation is working to two-loop
orders. The term (2.28) is accumulating the factor $(-\Box+m^{2})$
and thus could give rise to some new features. We want to confirm that
\begin{eqnarray}
\varphi_{3}(t,x)&\sim& (-\Box+m^{2})\int_{0}^{t}ds\int d^{4}y K(t-s,x-y)
\varphi_{2}(s,y)\varphi^{2}_{0}(s,y)\\
&\sim&(-\Box+m^{2})\int_{0}^{t}ds\int d^{4}y K(t-s,x-y)
(-\Box+m^{2})^{2}\varphi_{0}(s,y)g(s)\varphi^{2}_{0}(s,y)\nonumber
\end{eqnarray}
is convergent.
We thus examine the contribution arising from $\langle(-\Box+m^{2})^{2}\varphi_{0}(s,y)g(s)\\
\times \varphi_{0}(s,y)\rangle$,
\begin{eqnarray}
J(t)&=&\int_{0}^{t}ds\int [dq](q^{2}+m^{2})e^{-2d(q)s}g(s)
\end{eqnarray}
which is a 3-loop effect and convergent with $J(0)=0$. See (A.10) in Appendix. We thus have
\begin{eqnarray}
\varphi_{3}(t,x)
&\sim&(-\Box+m^{2})\varphi_{0}(t,x)J(t).
\end{eqnarray}

Our analysis suggests that the probe variable $\Phi(t,x)$ is well defined in the present scheme to all orders in perturbation theory. The most divergent terms arise from the tadpole-type contraction of 
$\langle \phi(x)\phi(y)\rangle$ in the probe variable, which contain loops of $\phi(x)$ detached from the proper part of Green's functions in the original $\lambda\phi^{4}$ theory. Since we are working in perturbation theory, the higher order corrections in the original theory
modify the behavior of those $\langle \phi(x)\phi(y)\rangle$ by some logarithmic factors up to any finite order in perturbation theory. Those logarithmic corrections to the propagator do not alter our analyses performed so far. 

We also need to analyze the correlation functions of the probe variable, for example,
\begin{eqnarray}
\langle \Phi(t,x)\Phi(t,y)\rangle.
\end{eqnarray}
In the lowest tree level, the correlation function is well defined for $t>0$ as is shown by $\langle \varphi_{0}(t,x)\varphi_{0}(t,x)\rangle$ in  (A.2) in Appendix and $\int_{0}^{\infty} dt \langle \Phi(t,x)\Phi(t,x)\rangle$ is finite. 
In the one-loop order, we have two (tadpole-type) finite corrections to the probe variable $\Phi(t,x)$ itself as discussed in (2.21) as a one-loop correction to $\varphi_{1}(t,x)$, and we also have a conventional tadpole diagram for the dynamical variable $\phi(x)$ bridging the two probe variables, whose divergence is removed by the mass counter term. The correlation (2.34) is well-defined to one-loop order even at $x=y$ for $t>0$. 

In the next two-loop order (by recalling the expansion (2.9)), we have, for example,
\begin{eqnarray}
&&\langle \varphi_{1}(t,x)\varphi_{0}(t,y)\rangle\nonumber\\
&&=[-(-\Box+m^{2})_{x}]\int_{0}^{t}ds\int d^{4}z  K(t-s,x-z)\nonumber\\
&&\times \int d^{4}z_{1}d^{4}z_{2}d^{4}z_{3}K(s,z-z_{1})K(s,z-z_{2})K(s,z-z_{3})\nonumber\\
&&\times\int d^{4}z_{4} K(t,y-z_{4})
\langle\phi(z_{1})\phi(z_{2})\phi(z_{3})\phi(z_{4})\frac{\lambda}{4} \int d^{4}w\phi^{4}(w)\rangle,
\end{eqnarray}
which is confirmed to be finite.
A calculation which is closely related to this contribution is performed later in (3.20).

\section{D+1 Dimensional theory}

We here present a formal proof of the finiteness of our scheme to all orders in perturbation theory following the 
$D+1$ dimensional formulation in the case of pure Yang-Mills theory~\cite{luscher2}.
The starting action is 
\begin{eqnarray}
S=\int d^{4}x {\cal L}_{D}(x)+\int_{0}^{\infty} dt\int d^{4}x{\cal L}_{D+1}(t,x)
\end{eqnarray}
where
\begin{eqnarray}
{\cal L}_{D}(x)&=&-\frac{1}{2}\partial_{\mu}\phi(x)\partial_{\mu}\phi(x)-\frac{1}{2}m^{2}\phi^{2}(x)-\frac{1}{4}\lambda\phi^{4}(x) +{\cal L}_{counter},\nonumber\\
{\cal L}_{D+1}(t,x)&=&L(t,x)[\partial_{t}\Phi(t,x)+\hat{D}(\hat{D}\Phi(t,x)+\lambda\Phi^{3}(t,x))]
\end{eqnarray}
with $\hat{D}\equiv-\Box+m^{2}$ and the constraint 
\begin{eqnarray}
\Phi(0,x)=\phi(x),
\end{eqnarray}
but no constraint on $L(0,x)$.
The free propagators are written as 
\begin{eqnarray}
&&\langle \Phi(t,x) L(s,y)\rangle=\theta(t-s)K(t-s,x-y),\nonumber\\
&&\langle \Phi(t,x)\hat{D}_{y} L(s,y)\rangle=\theta(t-s)\hat{D}_{x} K(t-s,x-y),\nonumber\\
&&\langle L(t,x)\phi(y)\rangle=0,\nonumber\\
&&\langle L(t,x)L(s,y)\rangle=0,\nonumber\\
&&\langle \Phi(t,x)\phi(y)\rangle=\int d^{4}x^{\prime}K(t,x-x^{\prime})\langle T^{\star}\phi(x^{\prime})\phi(y)\rangle,\\
&&\langle \Phi(t,x)\Phi(s,y)\rangle=\int d^{4}x^{\prime}K(t,x-x^{\prime})\int d^{4}y^{\prime}K(s,y-y^{\prime})\langle T^{\star}\phi(x^{\prime})\phi(y^{\prime})\rangle,\nonumber
\end{eqnarray}
in addition to the 4-dimensional $\langle T^{\star}\phi(x)\phi(y)\rangle$ with
\begin{eqnarray}
K(t,x-y)
&=&\int \frac{d^{4}k}{(2\pi)^{4}}e^{-(k^{2}+m^{2})^{2}t+ik(x-y)},
\end{eqnarray}
and the interaction terms are,
\begin{eqnarray}
\int_{0}^{\infty} dt\int d^{4}x{\cal L}_{int}(t,x)&=&\int_{0}^{\infty} dt\int d^{4}x\lambda \hat{D}L(t,x)\Phi^{3}(t,x),\nonumber\\
\int d^{4}x{\cal L}_{int}(x)&=&\int d^{4}x[-\frac{\lambda}{4} \phi^{4}(x)+ {\cal L}_{c}(x)].
\end{eqnarray}
With the convention of no closed loops of the bulk propagator due to the factor $\theta(t-s)$ that includes~\cite{luscher2}
\begin{eqnarray}
\langle \Phi(t,x)L(t,x)\rangle=0
\end{eqnarray}
and also $\langle \Phi(t,x)\hat{D}_{x}L(t,x)\rangle=0$, which are valid in dimensional regularization, for example, and the propagator 
\begin{eqnarray}
\langle T^{\star}\Phi(t,x)\Phi(s,y)\rangle&=&\int d^{4}x^{\prime}K(t,x-x^{\prime})\int d^{4}y^{\prime}K(s,y-y^{\prime})\langle T^{\star}\phi(x^{\prime})\phi(y^{\prime})\rangle\nonumber\\
&=&\int [dk]e^{-d(k)(t+s)+ik(x-y)}\frac{1}{k^{2}+m^{2}}
\end{eqnarray}
which is finite even at $x=y$ for any $t+s>0$, the bulk field theory defined for $t>0$ is confirmed to be finite as in the case of pure Yang-Mills theory~\cite{luscher2}. Recall that 
\begin{eqnarray}
d(k)=(k^{2}+m^{2})^{2}.
\end{eqnarray}
The conventional 4-dimensional $\lambda\phi^{4}$ theory is also rendered finite by the ordinary renormalization procedure. 

We thus analyze the possible divergence located near the boundary  $t=0$. Near $t=0$ we have for (3.8)
\begin{eqnarray}
\langle \Phi(t,x)\Phi(t,x)\rangle\simeq \delta(t)\int[dk]\frac{1}{2d(k)}\frac{1}{k^{2}+m^{2}}
\end{eqnarray}
in the sense of distribution with a finite numerical factor (see also Ref.~\cite{luscher2}).
The estimate of (3.10) is based on the use of a smooth test function $f(t)$, and we have
\begin{eqnarray}
&&\int_{0}^{\infty}dt f(t)\int [dk]e^{-2d(k)t}\frac{1}{k^{2}+m^{2}}\nonumber\\
&=&\int  [dk]\int_{0}^{\infty}d\tau f(\frac{\tau}{2d(k)})e^{-\tau}\frac{1}{2d(k)}\frac{1}{k^{2}+m^{2}}\nonumber\\
&=&f(0)\int  [dk]\frac{1}{2d(k)}\frac{1}{k^{2}+m^{2}}+\ {\rm  more\ convergent\ terms},
\end{eqnarray}
by expanding $f(\frac{\tau}{2d(k)})$ around the origin. We thus set 
\begin{eqnarray}
\int [dk]e^{-2d(k)t}\frac{1}{k^{2}+m^{2}}\rightarrow \delta(t)\int  [dk]\frac{1}{2d(k)}\frac{1}{k^{2}+m^{2}},
\end{eqnarray}
and this estimate of the leading term (3.10)
is consistent with $\langle \Phi(0,x)\Phi(0,x)\rangle=\infty$ and $\int_{0}^{\infty}dt\langle \Phi(t,x)\Phi(t,x)\rangle=\int [dk]\frac{1}{2d(k)}\frac{1}{k^{2}+m^{2}}$ in (3.8).

Thus the one-loop correction to the bulk vertex
\begin{eqnarray}
&&3\lambda \int_{0}^{\infty} dt \hat{D}L(t,x)\langle\Phi(t,x)\Phi(t,x)\rangle\Phi(t,x)\nonumber\\
&&=3\lambda \hat{D}L(0,x)\Phi(0,x)\int[dk]\frac{1}{2d(k)}\frac{1}{k^{2}+m^{2}}
\end{eqnarray}
is finite, which is consistent with our analysis in 4-dimensional formulation
in Section 2 where a more explicit evaluation is possible. This term is logarithmically  divergent for the conventional choice $d(k)=k^{2}+m^{2}$ and it would require a local counter term,
\begin{eqnarray}
\hat{D}L(0,y)\Phi(0,y)
\end{eqnarray}
at the boundary; to be precise, a local counter term for $d(k)=k^{2}+m^{2}$ would be $L(0,y)\Phi(0,y)$. 

We next analyze
\begin{eqnarray}
&&3\lambda^{2}\int_{0}^{\infty} dt\int_{0}^{\infty} ds \int d^{4}x\hat{D}L(t,x)\langle \Phi^{3}(t,x)\hat{D}_{y}L(s,y)\Phi^{2}(s,y)\rangle \Phi(s,y)\nonumber\\
&&=3!3\lambda^{2}\int_{0}^{\infty} dt\int_{0}^{\infty} ds \int  d^{4}x\hat{D}L(t,x)\Phi(s,y)\nonumber\\
&&\times \langle \Phi(t,x)\hat{D}_{y}L(s,y)\rangle\langle \Phi(t,x)\Phi(s,y)\rangle\langle \Phi(t,x)\Phi(s,y)\rangle\nonumber\\
&&=3!3\lambda^{2}\int_{0}^{\infty} dt\int_{0}^{\infty} ds  \int [dp](2\pi)^{4}\delta(p-q_{1}-q_{2}-q_{3})e^{-ipy}\hat{D}L(t,p)\Phi(s,y)\nonumber\\
&&\times \int[dq_{1}]\theta(t-s)e^{-d(q_{1})(t-s)}(q^{2}_{1}+m^{2})\int[dq_{2}]e^{-d(q_{2})(t+s)}\frac{1}{q_{2}^{2}+m^{2}}\nonumber\\
&&\times\int[dq_{3}]e^{-d(q_{3})(t+s)}\frac{1}{q_{3}^{2}+m^{2}}\nonumber\\
&&=3!3\lambda^{2}\int_{0}^{\infty} dt\int_{0}^{\infty} ds \int [dp]e^{-ipy}\hat{D}L(t,p)\Phi(s,y)\int[dq_{1}]\theta(t-s)e^{-d(q_{1})(t-s)}\nonumber\\
&&\times (q^{2}_{1}+m^{2})\int[dq_{2}]e^{-\left(d(q_{2})+d(p-q_{1}-q_{2})\right)(t+s)}\frac{1}{q_{2}^{2}+m^{2}}\frac{1}{(p-q_{1}-q_{2})^{2}+m^{2}}.
\end{eqnarray}
To analyze singular terms in this expression,  we replace 
\begin{eqnarray}
&&\theta(t-s)e^{-d(q_{1})(t-s)}\rightarrow \delta(t-s)\frac{1}{d(q_{1})},\nonumber\\
&&e^{-\left(d(q_{2})+d(p-q_{1}-q_{2})\right)(t+s)}\rightarrow \delta(t+s)\frac{1}{d(q_{2})+d(p-q_{1}-q_{2})}
\end{eqnarray}
in the sense of distribution. Eq.(3.15) thus becomes
\begin{eqnarray}
&&\frac{3!3}{2}\lambda^{2}\int [dp]e^{-ipy}\hat{D}L(0,p)\Phi(0,y)\\
&&\times \int[dq_{1}][dq_{2}]\frac{(q^{2}_{1}+m^{2})}{d(q_{1})}\frac{1}{d(q_{2})+d(p-q_{1}-q_{2})}\frac{1}{q_{2}^{2}+m^{2}}\frac{1}{(p-q_{1}-q_{2})^{2}+m^{2}}\nonumber
\end{eqnarray}
which is convergent and does not require any local counter term. If one uses the conventional $d(q)=q^{2}+m^{2}$, this integral is logarithmically divergent and it would require a local counter term in (3.14).

One can confirm that the one-loop term
\begin{eqnarray}
3^{2}\lambda^{2}\int_{0}^{\infty} dt\int_{0}^{\infty} ds \int d^{4}x\hat{D}L(t,x)\Phi(t,x)\langle \Phi^{2}(t,x)\hat{D}_{y}L(s,y)\Phi(s,y)\rangle \Phi^{2}(s,y)
\end{eqnarray}
is finite even for the conventional choice $d(q)=q^{2}+m^{2}$ without $\hat{D}$.

To analyze the absence of an extra wave function renormalization of $\phi(x)$, we next examine the two-loop diagrams 
\begin{eqnarray}
\lambda^{2}\int_{0}^{\infty} dt \hat{D}L(t,x)\langle \Phi^{3}(t,x)\phi^{3}(y)\rangle \phi(y)
\end{eqnarray}
which are confirmed to be finite; the tadpole of $\Phi(t,x)$ is finite as was shown above and the divergence in the tadpole of $\phi(x)$ is cancelled by the local mass counter term in ${\cal L}_{counter}$.
We thus analyze a diagram, which is analogous to the two loop self-energy
correction in $\lambda\phi^{4}$ theory,
\begin{eqnarray}
&&\int dt d^{4}x \hat{D}L(t,x)\langle \Phi^{3}(t,x)\phi^{3}(y)\rangle \phi(y)\nonumber\\
&&=3!\int dt d^{4}x [dp][dp_{1}][dp_{2}][dp_{3}]\exp\{-i(p_{1}+p_{2}+p_{3})(x-y)+ipx\}\hat{D}L(t,p) \phi(y)\nonumber\\
&&\times\exp\{-\left(d(p_{1})+d(p_{2})+d(p_{3})\right)t\}\frac{1}{(p^{2}_{1}+m^{2})(p^{2}_{2}+m^{2})(p^{2}_{3}+m^{2})}\nonumber\\
&&=3!\int dt[dp][dp_{1}][dp_{2}]e^{-ipy}\hat{D}L(t,p) \phi(y)\exp\{-\left(d(p_{1})+d(p_{2})+d(p-p_{1}-p_{2})\right)t\}\nonumber\\
&&\times\frac{1}{(p^{2}_{1}+m^{2})(p^{2}_{2}+m^{2})((p-p_{1}-p_{2})^{2}+m^{2})}\nonumber\\
&&=3!\int dt[dp]e^{-ipy}\hat{D}L(t,p) \phi(y)\{\int[dp_{1}][dp_{2}]\frac{\delta(t)}{d(p_{1})+d(p_{2})+d(p-p_{1}-p_{2})}\nonumber\\
&&\times\frac{1}{(p^{2}_{1}+m^{2})(p^{2}_{2}+m^{2})((p-p_{1}-p_{2})^{2}+m^{2})}+ {\rm finite\ terms}\}.
\end{eqnarray}
In this last expression, the integral over $[dp_{1}][dp_{2}]$ would be logarithmically divergent if one chooses the conventional $d(p)=p^{2}+m^{2}$,
which would in turn require a local counter term of the form (3.14).
But in our case with $d(p)=(p^{2}+m^{2})^{2}$, the integral is convergent and we do not need any local counter term.  

It is important that we have always the combination $\hat{D}L(t,x)$ for the variable $L(t,x)$ with $\hat{D}=-\Box+m^{2}$ due to our Feynman rules. See also (2.15).
The possible new counter term with the smallest dimension at the boundary, which was absent in the starting theory, is written as  
\begin{eqnarray}
\lambda^{l}z_{l}\int d^{4}x \hat{D}L(0,x)\phi(x) \hspace{1cm}{\rm for}\ \ l\geq 1,
\end{eqnarray}
with a suitable constant $z_{l}$; we note the relation $\hat{D}L(0,x)\Phi(0,x)=\hat{D}L(0,x)\phi(x)$. But the dimension of the operator $\hat{D}L(0,x)\phi(x)$ is 6 since the dimension of $L(0,x)$ is 3, and thus the operator  is irrelevant in the context of 4-dimensional theory and no divergent coefficients as in (3.13), (3.17) and (3.20)~\footnote{It is interesting that the possible BRST invariant local counter term in~\cite{luscher2} is
\begin{eqnarray}
\left(D_{\mu}L_{\mu}(0,x)-\{d,\bar{d}\}(0,x)\right)\partial_{\nu}A^{R}_{\nu}(x)-\bar{d}(0,x)
\partial_{\nu}D_{\nu}c^{R}(x)\nonumber
\end{eqnarray}
which also has dimension 6 and irrelevant.}. 

We have no extra possible local counter terms with dimensions less than or equal to 4; the dimension 4 local counter terms $\Phi^{4}(0,x)$, $\Phi^{3}(0,x)\phi(x)$, $\Phi^{2}(0,x)\phi^{2}(x)$, $\Phi(0,x)\phi^{3}(x)$, $\Phi(0,x)\Box\Phi(0,x)$ and $\Phi(0,x)\Box\phi(x)$ and the dimension 2 local counter terms $\Phi^{2}(0,x)$ and $\Phi(0,x)\phi(x)$ are not generated  by the vanishing closed loops of directed bulk propagators $\langle \Phi(t,x) L(s,y)\rangle$ combined with $\langle L(t,x) L(s,y)\rangle=0$ and $\langle \phi(x)L(s,y)\rangle=0$,
 and the counter term such as $\Phi^{3}(0,x)=\phi^{3}$ is excluded by reflection symmetry $\phi\rightarrow -\phi,\ \Phi\rightarrow -\Phi$, and $L\rightarrow -L$ in the starting action $S$. The theory is thus finite without any extra counter term other than ${\cal L}_{counter}$ required for the original $\lambda\phi^{4}$ theory in $D=4$.   
  
This mechanism to avoid the extra wave function renormalization is different from the case of pure Yang-Mills theory where the wave function renormalization factor for 
the external legs of 4-dimensional gauge field $A_{\mu}(x)$ is cancelled by the quantum corrections to the bulk probe variable $B_{\mu}(t,x)$~\cite{luscher2}. In our case both  probe variable $\Phi(t,x)$ and dynamical variable $\phi(x)$ are  renormalized variables, and to be consistent with this assumption, no extra divergence is induced in the interaction of the probe and 4-dimensional dynamical variables.

\section{ Composite operators}

The simplest composite operator in $\lambda\phi^{4}$ theory is~\cite{monahan} 
\begin{eqnarray}
T^{\star}\phi(x)\phi(y).
\end{eqnarray}
A way to analyze this composite operator in the gradient flow scheme is to examine
\begin{eqnarray}
&&\Phi(t,x)\Phi(t,x)\nonumber\\
&&=[\varphi_{0}(t,x)+\lambda\varphi_{1}(t,x)+ ...][\varphi_{0}(t,x)+\lambda\varphi_{1}(t,x)+ ...]\nonumber\\
&&=\varphi_{0}(t,x)\varphi_{0}(t,x)+\lambda\varphi_{0}(t,x)\varphi_{1}(t,x)+\lambda\varphi_{1}(t,x)\varphi_{0}(t,x)+
..\nonumber\\
&&=\int d^{4}yd^{4}z K(t,x-y)K(t,x-z)\phi(y)\phi(z)\nonumber\\
&&-2\lambda\int d^{4}zK(t,x-z)(-\Box+m^{2})_{x}\int d^{4}y\int_{0}^{t}ds K(t-s,x-y)\nonumber\\
&&\times \int d^{4}z_{1}d^{4}z_{2}d^{4}z_{3}K(s,y-z_{1})K(s,y-z_{2})K(s,y-z_{3})\phi(z)\phi(z_{1})\phi(z_{2})\phi(z_{3})\nonumber\\
&&+... . 
\end{eqnarray}
In this case, we obtain the expression to the linear order in $\lambda$
\begin{eqnarray}
\langle\Phi(t,x)\Phi(t,x)\rangle &\simeq& \frac{1}{32\pi^{2}}[\sqrt{\frac{\pi}{2}}\frac{1}{\sqrt{t}}+m^{2}\ln(tm^{4})]\nonumber\\
&+&\frac{\lambda}{8}\frac{m^{2}}{(16\pi^{2})^{2}}\ln(m^{4}t)\left(\ln\frac{m^{2}}{4\pi\mu^{2}}+\gamma_{E}-1\right)\nonumber\\
&-&\frac{3\lambda}{2}\frac{1}{(16\pi^{2})^{2}}[\sqrt{\frac{\pi}{2}}\frac{1}{\sqrt{t}}+\frac{m^{2}}{2}\ln(tm^{4})] 
\end{eqnarray}
for $t\rightarrow 0$ but $t\neq 0$. Here $\mu$ stands for the renormalization point of the mass term in the dimensional regularization. This expression may be compared with $\langle E\rangle $ in~\cite{luscher1}.

For any finite $t$, $0< t<\infty$, the above operator $\Phi(t,x)\Phi(t,x)$ is  finite. This implies that a singular factor $Z(t)$
\begin{eqnarray}
\Phi(t,x)\Phi(t,x)\rightarrow Z(t)\phi(x)\phi(x),
\end{eqnarray}
appears for small $t\rightarrow 0$. In comparison, in the case of the operator $m^{2}_{0}\phi^{2}_{0}(x)$ in $\lambda\phi^{4}$ theory with dimensional regularization, we have $m^{2}_{0}=Z_{m}m^{2}/Z_{\phi}$ and $\phi_{0}(z)=\sqrt{Z_{\phi}}\phi(z)$, and the finite dimension 2 operator is written as $m^{2}_{0}\phi^{2}_{0}(x)=m^{2}Z_{m}\phi^{2}(x)$.

To see the appearance of the singular factor $Z(t)$, we examine a specific order $\lambda$ correction in the perturbative expansion of $\langle \Phi(t,x)\Phi(t,x)\phi(w_{1})\phi(w_{2})\rangle$,
\begin{eqnarray}
&& -\frac{\lambda}{4} \int d^{4}w\langle \Phi(t,x)\Phi(t,x)\phi(w)^{4}\phi(w_{1})\phi(w_{2})\rangle\nonumber\\
&&=- 3 \lambda\int d^{4}w\langle \Phi(t,x)\phi(w)\rangle\langle \Phi(t,x)\phi(w)\rangle\langle \phi^{2}(w)\phi(w_{1})\phi(w_{2})\rangle.\nonumber\\
\end{eqnarray}
We thus evaluate
\begin{eqnarray}
&&\int d^{4}x\langle \Phi(t,x)\phi(w)\rangle\langle \Phi(t,x)\phi(w)\rangle\nonumber\\
&&=\int \frac{d^{4}k}{(2\pi)^{4}}e^{-(k^{2}+m^{2})^{2}t}\frac{1}{k^{2}+m^{2}}e^{-(k^{2}+m^{2})^{2}t}\frac{1}{k^{2}+m^{2}}
\nonumber\\
&&=\int_{0}^{\infty} \frac{xdx}{16\pi^{2}}e^{-2(x+m^{2})^{2}t}\frac{1}{(x+m^{2})^{2}}
\nonumber\\
&&=\int_{m^{2}}^{\infty} \frac{(y-m^{2})dy}{8\pi^{2}}\int_{t}^{\infty} dte^{-2y^{2}t}
\nonumber\\
&&\simeq \frac{1}{8\pi^{2}}[\int_{m^{2}}^{\infty} ydy\int_{t}^{\infty} dte^{-2y^{2}t} -\int_{m^{2}}^{\infty}m^{2} dy\int_{0}^{\infty} dte^{-2y^{2}t}]
\nonumber\\
&&=\frac{1}{8\pi^{2}}[\int_{m^{2}}^{\infty} ydy\int_{t}^{\infty} dte^{-2y^{2}t} -\int_{m^{2}}^{\infty}\frac{m^{2}}{2y^{2}}dy]
\nonumber\\
&&=\frac{1}{8\pi^{2}}[\int_{t}^{\infty}\frac{1}{4t}e^{-2m^{4}t}dt -\frac{1}{2}]
\nonumber\\
&&=\frac{1}{32\pi^{2}}[-\ln (m^{4}t)+ {\rm const.}].
\end{eqnarray}
Thus 
\begin{eqnarray}
&&\int d^{4}x\langle \Phi(t,x)\Phi(t,x)\phi(w_{1})\phi(w_{2})\rangle\\
&&=\{1+\frac{3\lambda}{32\pi^{2}}[\ln (m^{4}t)+ {\rm const.}]\}\int d^{4}x\langle \phi^{2}(x)\phi(w_{1})\phi(w_{2})\rangle,\nonumber
\end{eqnarray}
for $t\rightarrow 0$. The second term in (4.2) gives a contribution  included in the constant term.  
In the operator language, 
\begin{eqnarray}
\Phi(t,x)\Phi(t,x)\sim \langle \Phi(x,t)\Phi(x,t)\rangle+
\{1+\frac{3\lambda}{32\pi^{2}}[\ln (m^{4}t)+ {\rm const.}]\} [\phi^{2}](x)+ ...\nonumber\\
\end{eqnarray}
for $t\rightarrow 0$ but $t\neq 0$ with $[\phi^{2}](x)=\phi^{2}(x)-\langle \phi^{2}(0) \rangle$. Here we used the result in (4.3) in the first term.

We note that the formal energy-momentum tensor constructed from the quantity such as $\partial_{\mu}\Phi(t,x)\partial_{\nu}\Phi(t,x)$, 
\begin{eqnarray}
T_{\mu\nu}(t,x)&=&-\partial_{\mu}\Phi(t,x)\partial_{\nu}\Phi(t,x)\\
&&+g_{\mu\nu}[\frac{1}{2}\partial_{\mu}\Phi(t,x)\partial_{\mu}\Phi(t,x)+\frac{1}{2}m^{2}\Phi^{2}(t,x)+\frac{1}{4}\lambda\Phi^{4}(t,x)],\nonumber
\end{eqnarray}
which is expected to be finite for $t > 0$, does not generate translation of $\Phi(t,x)$ for $t > 0$ in general, since the canonical commutator is not defined for the variable $\Phi(t,x)$. Symmetry properties such as the derivation of Ward-Takahashi identities are less transparent in terms of the flowed variable $\Phi(t,x)$ for $t>0$. 

We expect that  $T_{\mu\nu}(t,x)$ in (4.9), which is based on renormalized variables and finite, is reduced  for $t\rightarrow 0$ to 
\begin{eqnarray}
T_{\mu\nu}(t,x)&\rightarrow &-Z_{\phi}(t)\partial_{\mu}\phi(x)\partial_{\nu}\phi(x)
+g_{\mu\nu}[\frac{1}{2}Z_{\phi}(t)\partial_{\mu}\phi(x)\partial_{\mu}\phi(x)\nonumber\\
&&+\frac{1}{2}Z_{m}(t)m^{2}\phi^{2}(x)+\frac{1}{4}Z_{\lambda}(t)\lambda\phi^{4}(x)],
\end{eqnarray}
in analogy with (4.4) with factors $Z_{\phi}(t), \ \ Z_{m}(t), \ \ Z_{\lambda}(t)$ which are finite for $t>0$ but divergent for the limit $t\rightarrow 0$. This expectation is based on the fact that the original finite energy-momentum tensor is written as  
\begin{eqnarray}
T_{\mu\nu}(x)&=&-\partial_{\mu}\phi_{0}(x)\partial_{\nu}\phi_{0}(x)\nonumber\\
&&+g_{\mu\nu}[\frac{1}{2}\partial_{\mu}\phi_{0}(x)\partial_{\mu}\phi_{0}(x)+\frac{1}{2}m_{0}^{2}\phi_{0}^{2}(x)+\frac{1}{4}\lambda_{0}\phi_{0}^{4}(x)],\nonumber\\
&=&-Z_{\phi}\partial_{\mu}\phi(x)\partial_{\nu}\phi(x)\\
&&+g_{\mu\nu}[\frac{1}{2}Z_{\phi}\partial_{\mu}\phi(x)\partial_{\mu}\phi(x)+\frac{1}{2}Z_{m}m^{2}\phi^{2}(x)+\frac{1}{4}Z_{\lambda}\lambda\phi^{4}(x)],\nonumber
\end{eqnarray}
and we expect that $1/\sqrt{t}$ plays a role similar to the cut-off $\Lambda^{2}$ for $t\rightarrow 0$.
The finiteness of  (the connected components of) the original energy-momentum tensor is inferred from  the Ward-Takahashi identity which is based on the conservation condition $\partial^{\mu}T_{\mu\nu}(x)=0$,
\begin{eqnarray}
&&\partial^{\mu}_{x}\langle T^{\star} T_{\mu\nu}(x)\phi_{0}(x_{1})\phi_{0}(x_{2})\phi_{0}(x_{3}) ....\phi_{0}(x_{n})\rangle\nonumber\\
&=&\sum_{k=1}^{n}\delta(x-x_{k})\langle T^{\star}  \phi_{0}(x_{1})\phi_{0}(x_{2})\partial_{\nu}^{x_{k}}\phi_{0}(x_{k}) ....\phi_{0}(x_{n})\rangle
\end{eqnarray}
and $\phi_{0}(x_{k})=\sqrt{Z_{\phi}}\phi(x_{k})$ in 
the dimensional regularization which eliminates quadratic divergences. When one divides both sides by $(\sqrt{Z_{\phi}})^{n}$, the right-hand side of (4.12) becomes finite and thus the left-hand side is also finite; we here forgo a necessary refinement of this argument.

A numerical analysis of the energy-momentum tensor for $\lambda\phi^{4}$ theory at $D=3$ in the framework of the free Gaussian flow has been performed  in~\cite{ehret}. Recent extensive analyses of the energy-momentum tensor for Yang-Mills theory in the gradient flow scheme are found in~\cite{suzuki2, deldebbio, makino, patella}.

\section{Conclusion}
The present analysis shows that the basic idea of the gradient flow in a narrow sense is consistently applied to the simple but important $\lambda\phi^{4}$ theory in $D=4$ and thus not limited to pure Yang-Mills theory.
 A crucial property, which is missing in the present $\lambda\phi^{4}$ theory compared to pure Yang-Mills theory, is the local gauge invariance in the Yang-Mills gradient flow. This absence of gauge invariance in $\lambda\phi^{4}$ theory  allowed us to introduce a regularization analogous to  the higher derivative regularization into the gradient flow equation; the most singular diagrams with closed bulk propagators are absent in the present example, and this  helps the higher derivative regularization work.

\subsection*{Acknowledgments}

I thank H. Suzuki for numerous clarifying comments on the subject of
gradient flow. This work is supported in part by JSPS KAKENHI (Grant No. 25400415). 

\appendix
\section{Some details of sample calculations}
We here summarize some details of sample calculations in Section 2.

We start with (2.26)
\begin{eqnarray}
\int_{0}^{t}ds\int d^{4}y K(t-s,x-y)[(-\Box_{y}+m^{2})\varphi_{0}(s,y)f(s)]\varphi_{0}(s,y)^{2}.
\end{eqnarray}
The contraction of $\langle \varphi_{0}(s,y)^{2}\rangle$ gives
\begin{eqnarray}
&&\langle \varphi_{0}(s,y)\varphi_{0}(s,y)\rangle\nonumber\\
&=&\int d^{4}z_{1}d^{4}z_{2}K(s,y-z_{1})]K(s,y-z_{2})\langle\phi(z_{1})\phi(z_{2})\rangle\nonumber\\
&=&\int [dk]\int d^{4}z_{1}d^{4}z_{2}\int[dp_{1}][dp_{2}]e^{-d(p_{1})s+ip_{1}(y-z_{1})}
e^{-d(p_{2})s+ip_{2}(y-z_{2})}\nonumber\\
&&\times \frac{e^{ik(z_{1}-z_{2})}}{k^{2}+m^{2}}\nonumber\\
&=&\int [dk]\frac{e^{-2d(k)s}}{k^{2}+m^{2}}.
\end{eqnarray}
We thus have
\begin{eqnarray}
&&\int_{0}^{t}ds\int d^{4}y\int [dp] e^{-d(p)(t-s)+ip(x-y)}[(q^{2}+m^{2})\int d^{4}z\int [dq]e^{-d(q)s+iq(y-z)}\phi(z)f(s)]\nonumber\\
&&\times\int [dk]\frac{e^{-2d(k)s}}{k^{2}+m^{2}}\nonumber\\
&=&\int_{0}^{t}ds\int [dp] e^{-d(p)(t-s)+ipx}[(p^{2}+m^{2})e^{-d(p)s}\phi(p)f(s)]\int [dk]\frac{e^{-2d(k)s}}{k^{2}+m^{2}}\nonumber\\
&=&(-\Box+m^{2})\varphi_{1}(t,x)\int_{0}^{t}dsf(s)\int [dk]\frac{e^{-2d(k)s}}{k^{2}+m^{2}}.
\end{eqnarray}
The crucial factor is, using the explicit formula for $f(s)$ in (2.22),
\begin{eqnarray}
g(t)&\equiv&\int [dk^{\prime}]\int_{0}^{t}ds \frac{e^{-2d(k^{\prime})s}}{(k^{\prime})^{2}+m^{2}}\int [dk]\frac{1-e^{-2d(k)s}}{2d(k)}\frac{1}{k^{2}+m^{2}}\nonumber\\
&=&\int [dk^{\prime}][dk][\frac{1-e^{-2d(k^{\prime})t}}{2d(k^{\prime})}-\frac{1-e^{-2(d(k^{\prime})+d(k))t}}{2(d(k^{\prime})+d(k))}]\frac{1}{2d(k)(k^{2}+m^{2})((k^{\prime})^{2}+m^{2})}\nonumber\\
&=&\int [dk^{\prime}][dk][\frac{-e^{-2d(k^{\prime})t}}{2d(k^{\prime})}+\frac{e^{-2(d(k^{\prime})+d(k))t}}{2(d(k^{\prime})+d(k))}]\frac{1}{2d(k)(k^{2}+m^{2})((k^{\prime})+m^{2})}\nonumber\\
&&+\frac{1}{4}\int [dk^{\prime}][dk][\frac{1}{d(k^{\prime})(d(k^{\prime})+d(k))}]\frac{1}{(k^{2}+m^{2})((k^{\prime})^{2}+m^{2})}
\end{eqnarray}
which is obviously convergent and $g(0)=0$. This result is used in (2.28).

When one studies the contraction of $[(-\Box_{y}+m^{2})\varphi_{0}(s,y)]\varphi_{0}(s,y)$ in (2.26), one examines 
\begin{eqnarray}
&&\langle[(-\Box_{y}+m^{2})\varphi_{0}(s,y)]\varphi_{0}(s,y)\rangle\nonumber\\
&=&\int d^{4}z_{1}d^{4}z_{2}[(-\Box_{y}+m^{2})K(s,y-z_{1})]K(s,y-z_{2})\langle\phi(z_{1})\phi(z_{2})\rangle\nonumber\\
&=&\int [dk]\int d^{4}z_{1}d^{4}z_{2}\int[dp_{1}][dp_{2}](p_{1}^{2}+m^{2})e^{-d(p_{1})s+ip_{1}(y-z_{1})}
e^{-d(p_{2})s+ip_{2}(y-z_{2})}\nonumber\\
&&\times \frac{e^{ik(z_{1}-z_{2})}}{k^{2}+m^{2}}\nonumber\\
&=&\int [dk]e^{-2d(k)s}.
\end{eqnarray}
The expression (2.26) thus becomes
\begin{eqnarray}
&&\int_{0}^{t}ds\int d^{4}y K(t-s,x-y)[(-\Box_{y}+m^{2})\varphi_{0}(s,y)f(s)]\varphi_{0}(s,y)^{2}\nonumber\\
&=&\int_{0}^{t}ds\int d^{4}y[dp] e^{-d(p)(t-s)+ip(x-y)}\int [dk]e^{-2d(k)s}f(s)\varphi_{0}(s,y)\nonumber\\
&=&\int_{0}^{t}ds\int d^{4}y[dp] e^{-d(p)(t-s)+ip(x-y)}\int [dk]e^{-2d(k)s}f(s)\int d^{4}z\int [dq]e^{-d(q)s+iq(y-z)}\phi(z)\nonumber\\
&=&\int_{0}^{t}ds[dp] e^{-d(p)(t-s)+ipx}\int [dk]e^{-2d(k)s}f(s)e^{-d(p)s}\phi(p)\nonumber\\
&=&\varphi_{0}(t,x)\int_{0}^{t}ds\int [dk]e^{-2d(k)s}f(s).
\end{eqnarray}
We thus examine
\begin{eqnarray}
h(t)&\equiv&\int_{0}^{t}ds\int [dk]e^{-2d(k)s}f(s)\nonumber\\
&=&\int [dk^{\prime}]\int_{0}^{t}ds e^{-2d(k^{\prime})s}\int [dk]\frac{1-e^{-2d(k)s}}{2d(k)}\frac{1}{k^{2}+m^{2}}\nonumber\\
&=&\int [dk^{\prime}][dk][\frac{1-e^{-2d(k^{\prime})t}}{2d(k^{\prime})}-\frac{1-e^{-2(d(k^{\prime})+d(k))t}}{2(d(k^{\prime})+d(k))}]\frac{1}{2d(k)(k^{2}+m^{2})}\nonumber\\
&=&\frac{1}{4}\int [dk^{\prime}][dk][\frac{-e^{-2d(k^{\prime})t}}{d(k^{\prime})}+\frac{e^{-2(d(k^{\prime})+d(k))t}}{(d(k^{\prime})+d(k))}]\frac{1}{d(k)(k^{2}+m^{2})}\nonumber\\
&&+\frac{1}{4}\int [dk^{\prime}][dk][\frac{1}{d(k^{\prime})(d(k^{\prime})+d(k))}]\frac{1}{(k^{2}+m^{2})}
\end{eqnarray}
By recalling that $d(k)=(k^{2}+m^{2})^{2}$, one can confirm that the last term in (A.7) is well-convergent. The convergence property is about the same as $f(t)$ in (2.22), and 
$h(0)=0$. This result is used in (2.30).

We finally want to confirm that the term in (2.31)
\begin{eqnarray}
\varphi_{3}(t,x)&\sim& (-\Box+m^{2})\int_{0}^{t}ds\int d^{4}y K(t-s,x-y)
\varphi_{2}(s,y)\varphi^{2}_{0}(s,y)\\
&\sim&(-\Box+m^{2})\int_{0}^{t}ds\int d^{4}y K(t-s,x-y)
(-\Box+m^{2})^{2}\varphi_{0}(s,y)g(s)\varphi^{2}_{0}(s,y)\nonumber
\end{eqnarray}
is convergent by contracting the terms
\begin{eqnarray}
\langle (-\Box+m^{2})^{2}\varphi_{0}(s,y)\varphi_{0}(s,y)\rangle
&=&\int [dq](q^{2}+m^{2})e^{-2d(q)s}.
\end{eqnarray}
We thus examine 
\begin{eqnarray}
\varphi_{3}(t,x)
&\sim&(-\Box+m^{2})\int_{0}^{t}ds\int [dp]e^{ipx}e^{-d(p)(t-s)}e^{-d(p)s}\phi(p)\int [dq](q^{2}+m^{2})e^{-2d(q)s}g(s)\nonumber\\
&\sim&(-\Box+m^{2})\Phi_{0}(t,x)\int_{0}^{t}ds\int [dq](q^{2}+m^{2})e^{-2d(q)s}g(s)\nonumber
\nonumber
\end{eqnarray}
and thus the combination
\begin{eqnarray}
J(t)&=&\int_{0}^{t}ds\int [dq](q^{2}+m^{2})e^{-2d(q)s}g(s)\nonumber\\
&=&\int_{0}^{t}ds\int [dq](q^{2}+m^{2})e^{-2d(q)s}\int [dk^{\prime}][dk][\frac{1-e^{-2d(k^{\prime})s}}{2d(k^{\prime})}-\frac{1-e^{-2(d(k^{\prime})+d(k))s}}{2(d(k^{\prime})+d(k))}]\nonumber\\
&&\times\frac{1}{2d(k)(k^{2}+m^{2})((k^{\prime})^{2}+m^{2})}\nonumber\\
&=&\frac{1}{8}\int [dq][dk^{\prime}][dk]\{[\frac{1-e^{-2d(q)t}}{d(q)d(k^{\prime})}-\frac{1-e^{-2(d(q)+d(k^{\prime}))t}}{d(k^{\prime})(d(q)+d(k^{\prime}))}]\nonumber\\
&&-[\frac{1-e^{-2d(q)t}}{d(q)(d(k)+d(k^{\prime}))}-\frac{1-e^{-2(d(q)+d(k)+d(k^{\prime}))t}}{(d(q)+d(k)+d(k^{\prime})(d(k)+d(k^{\prime})}]\}\nonumber\\
&&\times\frac{(q^{2}+m^{2})}{d(k)(k^{2}+m^{2})((k^{\prime})^{2}+m^{2})}
\end{eqnarray}
and the possible divergent terms are
\begin{eqnarray}
&&\int [dq][dk^{\prime}][dk]\{[\frac{1}{d(q)d(k^{\prime})}-\frac{1}{d(k^{\prime})
(d(q)+d(k^{\prime}))}]\nonumber\\
&&-[\frac{1}{d(q)(d(k)+d(k^{\prime}))}-\frac{1}{(d(q)+d(k)+d(k^{\prime}))(d(k)+d(k^{\prime})}]\}\nonumber\\
&&\times\frac{(q^{2}+m^{2})}{d(k)(k^{2}+m^{2})((k^{\prime})^{2}+m^{2})}\nonumber\\
&=&\int [dq][dk^{\prime}][dk]\frac{1}{(d(q)+d(k^{\prime}))(d(q)+d(k)+d(k^{\prime}))}\nonumber\\
&&\times\frac{1}{(q^{2}+m^{2})(k^{2}+m^{2})((k^{\prime})^{2}+m^{2})}\nonumber\\
\end{eqnarray}
which is a 3-loop effect and convergent, and one obtains $J(0)=0$. 
This result is used in (2.33).

\end{document}